\documentclass[11pt,twoside]{article}
\usepackage{./asp2014}

\aspSuppressVolSlug
\resetcounters

\begin{document}

\title{TOPCAT's TAP Client}
\author{M.~B.~Taylor}
\affil{H.~H.~Wills Physics Laboratory, University of Bristol, U.K.
       \email{m.b.taylor@bristol.ac.uk}}

% This section is for ADS Processing.  There must be one line per author.
\paperauthor{M.~B.~Taylor}{m.b.taylor@bristol.ac.uk}{0000-0002-4209-1479}{University of Bristol}{School of Physics}{Bristol}{}{BS8 1TL}{U.K.}

\begin{abstract}
TAP, the Table Access Protocol,
is a Virtual Observatory (VO) protocol for
executing queries in remote relational databases using ADQL,
an SQL-like query language.
It is one of the most powerful components of the VO,
but also one of the most complex to use, with an extensive
stack of associated standards.

We present here recent improvements to the client and GUI
for interacting with TAP services
from the TOPCAT table analysis tool.
As well as managing query submission and result retrieval,
the GUI attempts to provide the user with as much help as possible
in locating services,
understanding service metadata and capabilities,
and constructing correct and useful ADQL queries.
The implementation and design are, unlike previous versions,
both usable and performant even for the largest TAP services.
\end{abstract}

\section{Introduction}

TAP, the Table Access Protocol \citep{2011arXiv1110.0497D},
is a Virtual Observatory
standard that allows clients to execute custom queries on remote databases
and retrieve the results for local use.
Although TAP allows in principle for a variety of query languages,
in this paper we restrict consideration to ADQL,
the Astronomical Data Query Language, which is the only
one mandated by TAP.
ADQL is essentially a standardised dialect of the SQL SELECT
statement,
and as such allows complex queries against a relational database,
including such functions as row selections, column combinations
and multi-table joins.
It also defines a number of optional geometrical functions to assist
in constraining results on the celestial sphere.
The flexibility afforded by this framework
permits very powerful queries to be made against potentially very
large and complex remote datasets.
In particular it is far more capable than the ``Simple'' VO protocols
for catalogue, image or spectral access that preceded it.

The price for this flexibility is however that TAP is a complex
beast, requiring use of a stack of related IVOA standards
including RegTAP, VOResource, VODataService, TAPRegExt,
VOTable, DALI and UWS, as well as TAP and ADQL themselves.
The hard part of the problem is not writing software to communicate
with external services using these standards, but presenting
an interface to the astronomy user that moderates the apparent
complexity.

TOPCAT\footnote{\url{http://www.starlink.ac.uk/topcat/}}
\citep{2005ASPC..347...29T} is a desktop GUI application,
widely used by astronomers,
for analysis of catalogues and other tabular data.
Since TAP services are a prime potential source of tables for
local analysis, it has an integrated TAP client as one of its
table load dialogue options.
This TAP load dialogue has been part of the application
since 2011, not long after the adoption of TAP
as an IVOA Recommendation, but it had a number of deficiencies,
so a major overhaul of the functionality has been introduced in
the recent version 4.3 (2015).
We describe the improved implementation and user interface in the
following sections.

The implementation described here is not tied to the TOPCAT application.
The libraries are available for external use,
either as a standalone TAP GUI client,
or for embedding in third-party java applications.

\section{Service Discovery}

In order to execute a TAP query, the user must first decide which
TAP service to use; at time of writing around 100
are listed in the IVOA Registry, and this number is expected to rise.
Different services have different data holdings,
from large general services such as TAPVizieR
\citep{2013ASPC..475..227L} with 30\,000 data tables,
to much more focussed ones like the Chandra Source Catalogue at Harvard
with only 3.

Typically, the astronomer knows which data sets she wants to use
(e.g.\ WISE or CALIFA) rather than the name or location of the service
hosting them (e.g.\ HEASARC or GAVO DC).
When selecting a TAP service to query therefore, it is important to be
able to locate services by searching against {\em table\/} metadata
such as table name and description,
rather than {\em service\/} metadata, which may or may not
contain table-level detail.

The standard way to locate VO services is to use the IVOA Registry,
a white pages for data services and other VO resources
\citep{2015A&C....10...88D}.
This can be used to discover TAP services, but unfortunately does
not currently contain sufficient table-level information to
identify them by table metadata as required.
To work round this limitation,
TOPCAT by default uses instead of the Registry a separate, non-standard,
database called GloTS.
GloTS, the Global TAP Schema, is maintained by the
German Astrophysical Virtual Observatory (GAVO),
being updated automatically by crawling registered TAP
services to retrieve the detailed table-level metadata they
declare, and exposing the results via a TAP service.
This enables just the kind of queries that TOPCAT requires
to support its table-oriented service discovery user interface.

Alternative approaches to providing table-level metadata within the
Registry are under investigation within the IVOA,
and may enable similar functionality by use of standardised
service interfaces in the future.
The service discovery functionality is implemented as a
pluggable layer within TOPCAT, to provide a platform for
experimenting with alternative ways to perform such searches.
A configuration option is present in the GUI to switch between
different service discovery backends, though this is only
currently useful as a platform for prototyping registry
experiments.

\section{Metadata Acquisition and Display}

Having selected a TAP service to use, the user needs to understand
exactly what is provided by the service in order to be able to
construct useful queries, and so TOPCAT has to make this information
available in the GUI.
The most important items constituting this service metadata are the
descriptions of available tables, and the descriptions
of columns in each table.
Although this information may be available from the Registry or GloTS,
the most reliable way to obtain it is from the service itself.
TAP defines two basic ways for services to provide this self-description:
as an XML document served from the {\tt /tables} endpoint and
as the content of some system-level tables within the
exposed database itself, in the reserved TAP\_SCHEMA namespace.

For most services, it is straightforward and efficient
to acquire this metadata
by downloading it in one go when the service is first contacted.
It can then be stored wholesale in the client and presented to
the user as required; for instance when a table is selected in the GUI,
names and descriptions for the columns of that table can be displayed.
However, for the small number of services which serve very many tables,
this approach is no longer practical.
The metadata for TAPVizieR's 30\,000 tables amounts to around
100\,Mbyte, only a tiny proportion of which a user will want
to examine in a single session, so downloading the whole thing
pre-emptively is not desirable.

To address this, TOPCAT uses a pluggable metadata acquisition layer,
with different backend strategies for different services.
By default, an adaptive strategy is used:
if there are fewer than 5000 columns in total, the metadata is
downloaded in one go from the XML document,
but if there are more, then TAP\_SCHEMA queries are made
in the first instance to acquire just a list of tables,
and the more bulky column metadata is retrieved as required using
subsequent per-table TAP\_SCHEMA queries only for those tables
the user expresses an interest in.
Again, there are expert options in the GUI for switching
between metadata acquisition strategies as required, but
normal users are not expected to need to be aware of these.

Having acquired the service metadata, the application must make it
available to the user through the GUI.
This is done using the combination of a selectable
tree of tables, with an adjacent panel providing more detail on
the currently selected one.
Since much information is available and screen real estate
is at a premium, the detail panel contains
a number of tabs describing different aspects of the selected
table: schema, table name/description, column list, and
foreign key information.

Scalability is an issue for GUI usability as well as
data transfer bandwidth.
With potentially thousands of tables from which to choose, browsing a
scrollable list is not a useful interface, especially when
tables have unintuitive names.
A text entry field is therefore provided to restrict the
content of the currently displayed list of tables to those
whose name and/or description matches one or more given search terms.
The list is filtered instantly as search terms are typed.

\section{Query Preparation}

Once in possession of
the available information about the currently targeted service,
the user has to assemble the text of an ADQL SELECT statement
specifying the desired database operation.
Most astronomers are not, at least initially, fluent in ADQL or SQL,
so need some assistance with the syntax.
One possible approach
is to provide a graphical query builder that
constructs a SELECT statement from a series of GUI interactions
(e.g.\ selecting tables, columns and comparison operations
from drop-down menus).
That can be effective for simple queries, but is difficult to
generalise to more sophisticated or flexible operations.
TOPCAT instead takes the approach of providing a libary of
example queries that a user can use, edit, adapt and learn from.

These examples are available from a menu that fills in the ADQL
text ready to submit, and fall into three categories.
{\em Standard\/} examples use standard TAP features,
and are constructed by TOPCAT with reference to metadata retrieved
from the current service, so can be used as-is to make working
(though not necessarily useful) queries on the database at hand.
{\em Data Model-Specific\/} examples consist of static pre-written
queries specific to particular data models for which the current
service declares support, if any; for instance if a service
declares support for the well-known ObsTAP data model, implying that the
{\tt ivoa.obscore} table is present with a well-known column structure,
then TOPCAT will offer a list of queries that make sense for that table.
{\em Service-Provided\/} examples consist of ADQL text retrieved from
the service itself in a standard format via the {\tt /examples} endpoint,
and can thus provide ready-to-run queries exploiting the
particular structure and capabilities of the current database.
This final category relies on recent enhancements to the TAP protocol
stack, for which the details are still under discussion,
but presents a powerful way for data providers to
assist end users in making best use of the archived data.

Finally, a {\em Hints\/} tab alongside the metadata display
provides a very basic ``cheat sheet'' with reminders about SELECT
statement syntax and pointers to a few external ADQL resources.

When actually assembling the ADQL query for submission, the user
types into a text entry panel.
Queries are validated as they are entered, with reference to
ADQL syntax and service-specific information such as the list
of available tables, columns and user-defined functions,
so that syntax errors can be highlighted.
Other features of the text entry panel include undo/redo,
multiple tabs, and limited support for pasting in column and table
names selected from the GUI.

\section{Conclusions}

It is hoped that TOPCAT's enhanced TAP user interface,
alongside parallel developments in other available TAP clients,
evolution of associated standards,
and continuing improvements in service implementations,
will lead to more widespread use of TAP
for effective exploitation
of the vast and increasing
amount of astronomical data
which is exposed using this protocol.

\acknowledgements This work has benefitted from assistance from many members of the TAP/IVOA community; special thanks are due to Markus Demleitner (GloTS \&\ expert on all things TAP) and Gr\'{e}gory Mantelet (ADQL parser library), both at ARI Heidelberg.  Some of the ideas here were inspired by other TAP clients, including Seleste \citep{2013AAS...22124025V} and TAPHandle \citep{2014ASPC..485...15M}.

\bibliography{P107}

\end{document}